\documentclass[sigconf]{acmart}

\AtBeginDocument{%
  \providecommand\BibTeX{{%
    \normalfont B\kern-0.5em{\scshape i\kern-0.25em b}\kern-0.8em\TeX}}}

\usepackage{enumitem}
\setlist[itemize]{noitemsep, topsep=0pt}

\usepackage{rotating}
\usepackage{booktabs}
\usepackage{multirow}
\usepackage{graphicx}
\usepackage{color}

\usepackage{array}
\usepackage{needspace}
\usepackage{multirow}

\usepackage{nth}
\usepackage{longtable}
\usepackage{supertabular}

\usepackage{enumitem}
\setlist{nolistsep,leftmargin=*}

\usepackage{pdfpages}

\newcommand{\xhdr}[1]{\vspace{1.7mm}\noindent{{\bf #1.}}}

\newcommand{\squeezetiny}[1]{\textls[-5]{#1}}
\newcommand{\squeeze}[1]{\textls[-10]{#1}}
\newcommand{\squeezesome}[1]{\textls[-15]{#1}}

\newcommand{\onlineappendix}[1]{Appendix~#1}
\newcommand{\irb}{Our study was approved by the Microsoft Research Institutional Review Board (IRB).}
\renewcommand\enlargethispage[1]{}

\newcommand{\addacmfooter}{
    \renewcommand\footnotetextcopyrightpermission[1]{}
    \settopmatter{printacmref=false}
    \pagestyle{plain}
    \fancyfoot{}
}
\newcommand{\setacmlicense}{}

\addacmfooter

\begin{document}

\setacmlicense

\title[Population-Scale Study of Human Needs During the COVID-19 Pandemic]{Population-Scale Study of Human Needs\\During the COVID-19 Pandemic: Analysis and Implications}


\author{Jina Suh}
\email{jinasuh@cs.washington.edu}
\affiliation{
  \institution{University of Washington}
}
\affiliation{
  \institution{Microsoft Research}
}

\author{Eric Horvitz, Ryen W. White}
\email{{horvitz,ryenw}@microsoft.com}
\affiliation{
  \institution{Microsoft Research}
}


\author{Tim Althoff}
\email{althoff@cs.washington.edu}
\affiliation{
  \institution{University of Washington}
}
\renewcommand{\shortauthors}{Jina Suh et al.}


\newcommand{\tabdataset}{
\begin{table}[]
    \small
    \centering
    \begin{tabular}{lr}
    \toprule
    Observation period & ~14 months (Jan 1-Aug 2 in 2019-2020) \\
    \# of queries & 35,650,687,581 \\
    \# of human need queries & 3,250,228,644 \\
    \# of days & 428 \\
    \# of ZIP codes & 36,667 \\
    \bottomrule
    \end{tabular}%
    \vspace{1pt}
    \captionsetup{width=\linewidth, labelfont=bf,textfont={bf},font=small,aboveskip=0pt, belowskip=0pt, singlelinecheck=false}
    \caption{Descriptive statistics for our web search dataset.}
    \label{tab:dataset}
    \vspace{-14pt}
\end{table}
}

\newcommand{\process}{
    \begin{figure*}[t!]
    \vspace{-1\baselineskip}
    \includegraphics[width=0.9\textwidth]{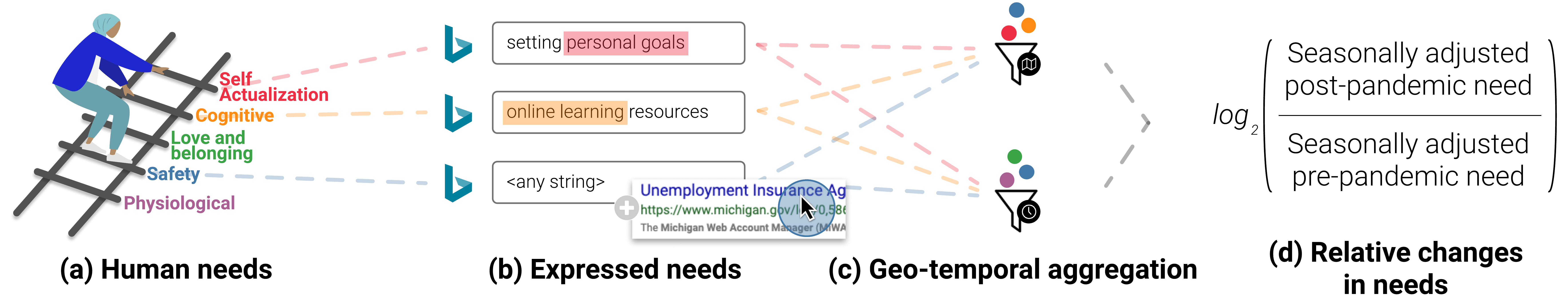}
    \vspace{0.5pt}
    \captionsetup{width=\linewidth, labelfont=bf,textfont={bf},font=small,aboveskip=0pt, belowskip=0pt, singlelinecheck=false}
    \caption{Illustration of human needs detection framework. (a) Human needs are represented by a ladder according to Maslow's hierarchy of needs to indicate than a person may have multiple needs simultaneously. (b) These needs are expressed through search interactions, which can be categorized through keyword matches and/or subsequent clicks into relevant search result pages. (c) Each search behavior is then aggregated across categories of human needs, time, or geography. (d) To quantify changes in needs, aggregated needs are compared between pre-pandemic and pandemic periods while adjusting for seasonal and query volume variation.}
    \Description{Illustration of human needs detection framework. (a) Human needs are represented by a ladder according to Maslow's hierarchy of needs to indicate than a person may have multiple needs simultaneously. (b) These needs are expressed through search interactions, which can be categorized through keyword matches and/or subsequent clicks into relevant search result pages. (c) Each search behavior is then aggregated across categories of human needs, time, or geography. (d) To quantify changes in needs, aggregated needs are compared between pre-pandemic and pandemic periods while adjusting for seasonal and query volume variation.}
    \label{fig:process}
    \vspace{-14pt}
    \end{figure*}
}

\newcommand{\annotateedtrend}{
    \begin{figure}[t]
    \vspace{-1\baselineskip}
    \includegraphics[width=0.95\linewidth]{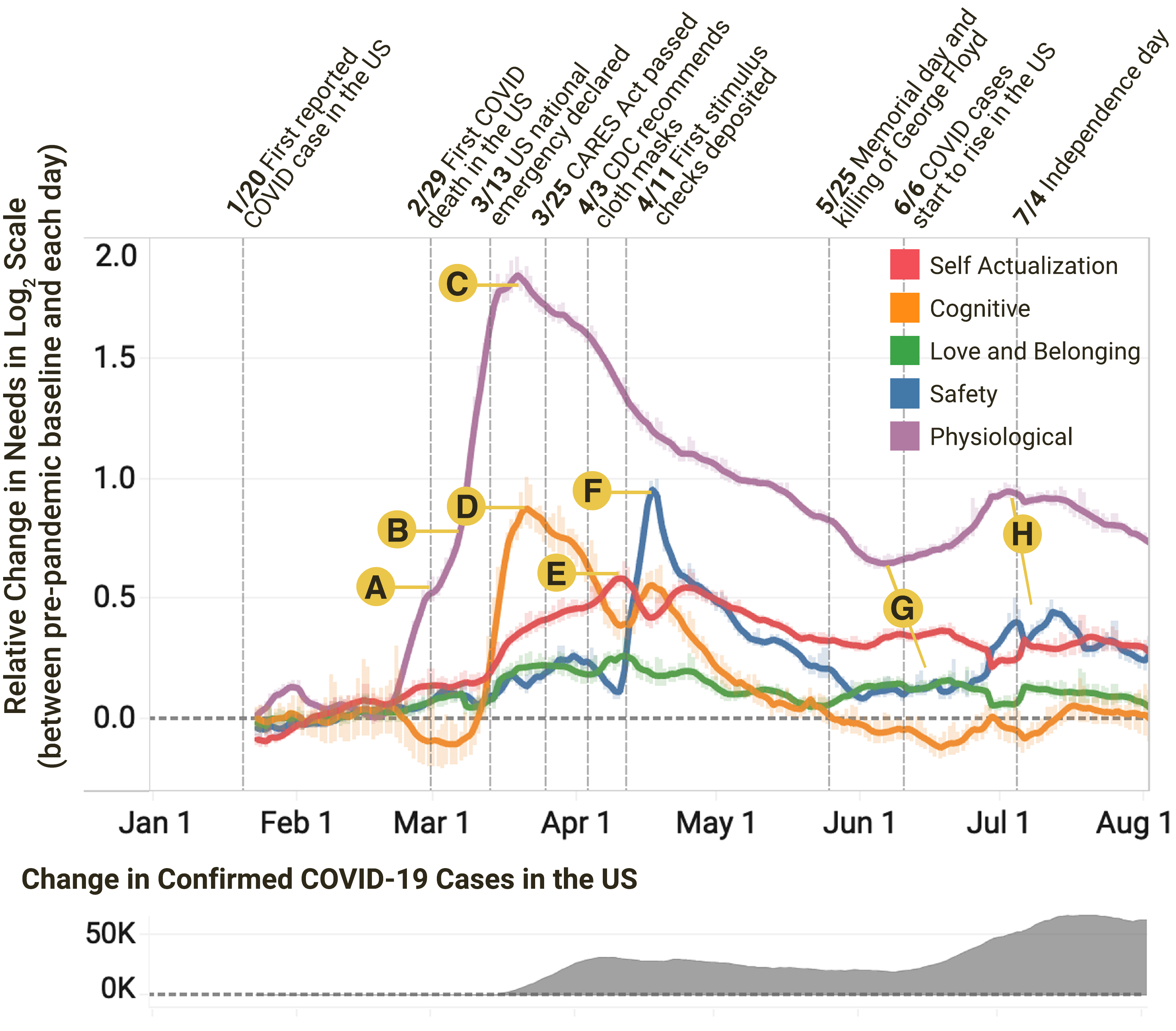}
    \vspace{0.5pt}
    \captionsetup{width=\linewidth, labelfont=bf,textfont={bf},font=small,aboveskip=0pt, belowskip=0pt, singlelinecheck=false}
    \caption{Daily relative changes in needs for each human need category throughout the pandemic. Relevant events are annotated vertically at the time of occurrence. Changes in confirmed cases in the US from `Bing COVID-19 Tracker' (\url{http://www.bing.com/covid}) are displayed below. See Sect.~\ref{sec:tempchange} for discussion on A-H.}
    \Description{Daily relative changes in needs for each human need category throughout the pandemic. Relevant events are annotated vertically at the time of occurrence. Changes in confirmed cases in the US from `Bing COVID-19 Tracker' (\url{http://www.bing.com/covid}) are displayed below. See Sect.~\ref{sec:tempchange} for discussion on A-H.}
    \label{fig:annotatedtrend}
    \vspace{-14pt}
    \end{figure}
}

\newcommand{\tabtopchange}{
\begin{table}[]
    \small
    \centering
    \vspace{-1\baselineskip}
    \begin{tabular}{llllr}
    \toprule
    \textbf{Need} & \textbf{Need Subcategory} & \textbf{$C_{\text{mean}}$} & \textbf{$C_{\text{max}}$} & \textbf{$2^{C_{\text{max}}}$-$1$}\\
    \midrule
    Phys & Toilet paper purchase & 6.11$\pm$0.08 & 7.00 & 12691.1\% \\
    Safe & Stimulus related queries & 5.69$\pm$0.03 & 8.17 & 28601.9\% \\
    Safe & Unemploy. related queries & 4.88$\pm$0.02 & 5.72 & 5156.4\% \\
    Safe & State unemploy. site visits & 4.81$\pm$0.04 & 6.26 & 7585.1\% \\
    Safe & COVID-19 prot. purchase & 4.67$\pm$0.03 & 5.57 & 4634.0\% \\
    Phys & Health meas. equip. purchase & 3.43$\pm$0.04 & 4.56 & 2257.2\% \\
    Phys & Health cond. related queries & 3.12$\pm$0.01 & 3.54 & 1065.6\% \\
    Phys & Food assist. related queries & 2.62$\pm$0.03 & 3.12 & 771.8\% \\
    Phys & Grocery related queries & 2.05$\pm$0.01 & 2.54 & 480.8\% \\
    Phys & Food delivery  queries & 1.84$\pm$0.02 & 2.26 & 379.7\% \\
    L\&B & Online social activity queries & 1.77$\pm$0.09 & 2.98 & 688.4\% \\
    Phys & Food delivery site visits & 1.7$\pm$0.01 & 2.26 & 379.3\% \\
    \bottomrule
    \end{tabular}%
    \vspace{1pt}
    \captionsetup{width=\linewidth, labelfont=bf,textfont={bf},font=small,aboveskip=0pt, belowskip=0pt, singlelinecheck=false}
    \caption{Top 12 need subcategories with the largest \emph{increase} in mean relative change in need within the initial 4 weeks of the pandemic with 95\% confidence intervals, maximum relative change in the dataset, and maximum percent change.}
    \label{tab:topchange}
    \vspace{-8pt}
\end{table}
}

\newcommand{\subneeds}{
    \begin{figure}[t]
    \vspace{-1\baselineskip}
    \includegraphics[width=0.9\linewidth]{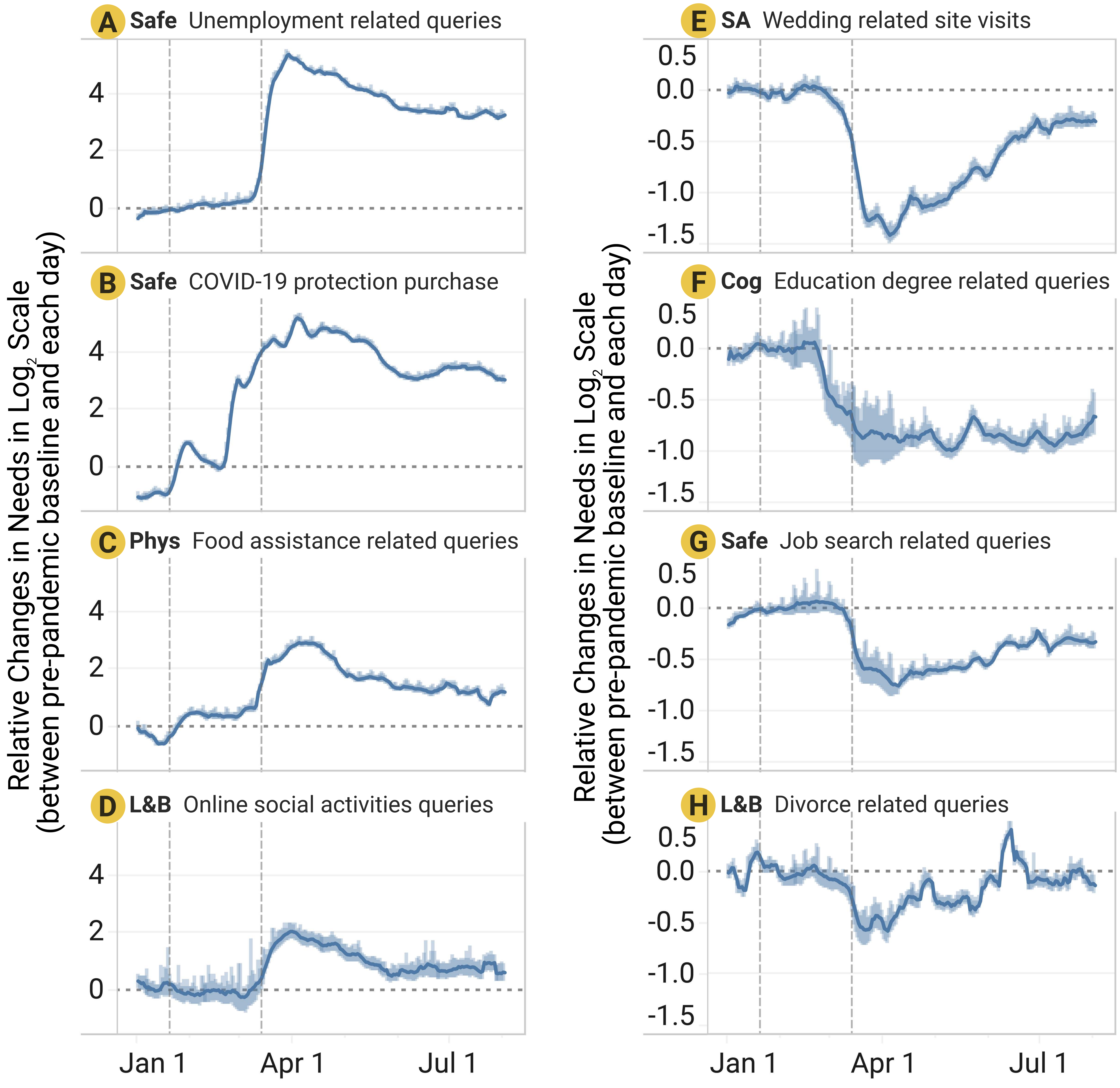}
    \vspace{0.5pt}
    \captionsetup{width=\linewidth, labelfont=bf,textfont={bf},font=small,aboveskip=0pt, belowskip=0pt, singlelinecheck=false}
    \caption{Daily relative changes in select needs with the largest increase (A-D) and the largest decrease (E-H) in need. Vertical bars denote the first reported US COVID case (Jan 20) and the US national emergency declaration (Mar 13).}
    \Description{Daily relative changes in select needs with the largest increase (A-D) and the largest decrease (E-H) in need. Vertical bars denote the first reported US COVID case (Jan 20) and the US national emergency declaration (Mar 13).}
    \label{fig:subneeds}
    \vspace{-14pt}
    \end{figure}
}

\newcommand{\tabbottomchange}{
\begin{table}[]
    \small
    \centering
    \begin{tabular}{lllll}
    \toprule
    \textbf{Need} & \textbf{Need Subcategory} & \textbf{$C_{\text{mean}}$} & \textbf{$C_{\text{min}}$} & \textbf{$2^{C_{\text{min}}}$-$1$}\\
    \midrule
    SA & Wedding related purchase & -1.49$\pm$0.03 & -1.76 & -70.4\% \\
    SA & Wedding site visits & -1.25$\pm$0.02 & -1.56 & -66.2\% \\
    Cog & Edu. degree related queries & -0.87$\pm$0.06 & -1.09 & -53.1\% \\
    Safe & Housing related queries & -0.71$\pm$0.05 & -1.09 & -53.2\% \\
    Safe & Job search related queries & -0.65$\pm$0.03 & -0.91 & -46.7\% \\
    Safe & Job search site visits & -0.61$\pm$0.02 & -0.95 & -48.1\% \\
    Phys & Apparel purchase & -0.60$\pm$0.01 & -0.84 & -44.1\% \\
    SA & Outdoor related queries & -0.59$\pm$0.01 & -1.07 & -52.3\% \\
    SA & Life goal related queries & -0.57$\pm$0.09 & -1.23 & -57.2\% \\
    Safe & Domestic violence queries & -0.54$\pm$0.04 & -0.97 & -49.0\% \\
    Safe & Rental related queries & -0.53$\pm$0.06 & -0.82 & -43.5\% \\
    L\&B & Divorce related queries & -0.49$\pm$0.03 & -0.93 & -47.3\% \\
    \bottomrule
    \end{tabular}%
    \vspace{1pt}
    \captionsetup{width=\linewidth, labelfont=bf,textfont={bf},font=small,aboveskip=0pt, belowskip=0pt, singlelinecheck=false}
    \caption{Top 12 need subcategories with the largest \emph{decrease} in mean relative change in need within the initial 4 weeks of the pandemic with 95\% confidence intervals, minimum relative change in the dataset, and minimum percent change.}
    \label{tab:bottomchange}
    \vspace{-14pt}
\end{table}
}

\newcommand{\stateshelter}{
    \begin{figure}[t]
    \vspace{-1.2\baselineskip}
    \includegraphics[width=0.87\linewidth]{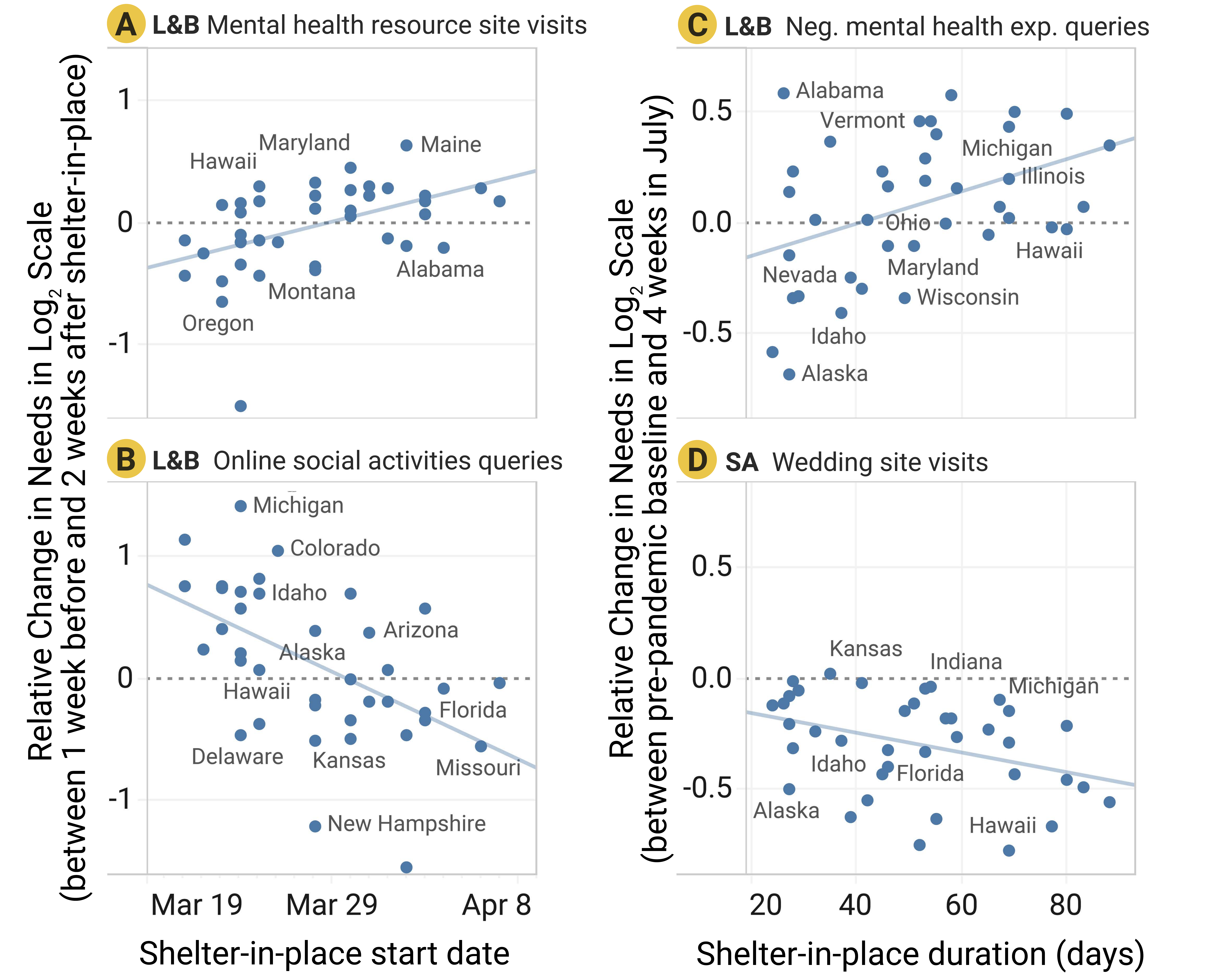}
    \vspace{1.5pt}
    \captionsetup{width=\linewidth, labelfont=bf,textfont={bf},font=small,aboveskip=0pt, belowskip=0pt, singlelinecheck=false}
    \caption{Relationships between shelter-in-place start date and \emph{short-term} relative changes in mental health and online social activities needs between 1 week before and 2 weeks after shelter-in-place mandates (A-B), and between shelter-in-place duration and \emph{long-term} relative changes in mental health and wedding needs between pre-pandemic baseline and 4 weeks in July (C-D).}
    \Description{Relationships between shelter-in-place start date and \emph{short-term} relative changes in mental health and online social activities needs between 1 week before and 2 weeks after shelter-in-place mandates (A-B), and between shelter-in-place duration and \emph{long-term} relative changes in mental health and wedding needs between pre-pandemic baseline and 4 weeks in July (C-D).}
    \label{fig:stateshelter}
    \vspace{-14pt}
    \end{figure}
}

\newcommand{\unemployment}{
    \begin{figure}[t]
    \vspace{-1\baselineskip}
    \includegraphics[width=0.9\linewidth]{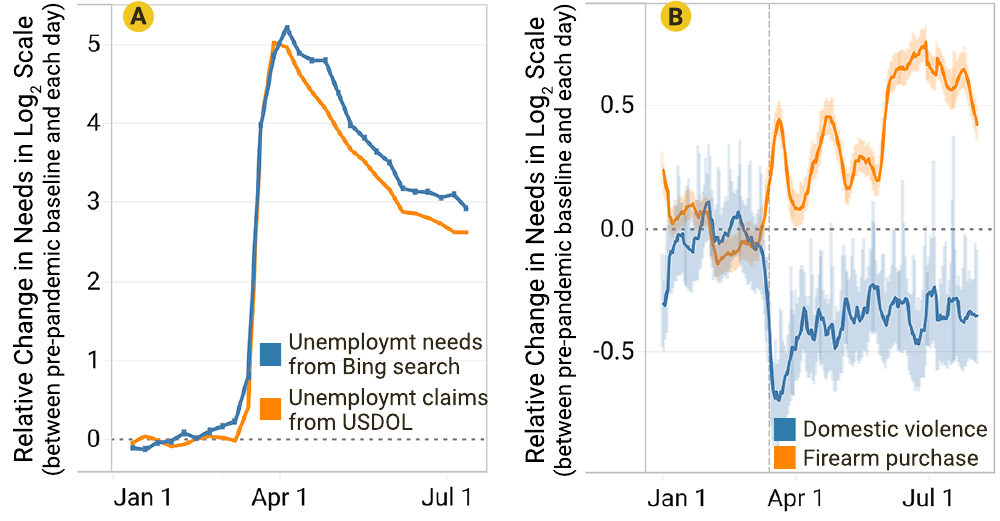}
    \vspace{1.5pt}
    \captionsetup{width=\linewidth, labelfont=bf,textfont={bf},font=small,aboveskip=0pt, belowskip=0pt, singlelinecheck=false}
    \caption{Weekly relative changes in unemployment needs from Bing search and in unemployment claims from the US Department of Labor compared to pre-pandemic baseline (A), and daily relative changes in domestic violence and firearm purchase needs compared to pre-pandemic baseline (B).}
    \Description{Weekly relative changes in unemployment needs from Bing search and in unemployment claims from the US Department of Labor compared to pre-pandemic baseline (A), and daily relative changes in domestic violence and firearm purchase needs compared to pre-pandemic baseline (B).}
    \label{fig:unemployment}
    \vspace{-14pt}
    \end{figure}
}

\newcommand{\suppdemographic}{
    \begin{table}[h]
    \small
    \begin{tabular}{lrr}
    \toprule
    & \textbf{Our Dataset} &  \textbf{Census}  \\
    \midrule
    Population & 3,072 & 2,775 \\
    Median   Income & \$54,231 & \$54,048 \\
    Median   Age & 41.8 & 41.9 \\
    \%   Race White & 0.87 & 0.88  \\
    \%   Male & 0.5 & 0.5 \\
    \%   HS Grad or Higher & 0.9 & 0.9 \\
    Gini   Index & 0.42 & 0.42 \\
    \%   Below Poverty Lvl. & 0.12 & 0.12 \\
    \%   Housing Owned & 0.76 & 0.76 \\
    \%   Has Internet & 0.77 & 0.76 \\
    \bottomrule
    \end{tabular}%
    \vspace{1pt}
    \captionsetup{width=\linewidth, labelfont=bf,textfont={bf},font=small,aboveskip=0pt, belowskip=0pt, singlelinecheck=false}
    \caption{Distribution of demographic variables (median) for US ZIP codes in our dataset compared to all of the available US ZIP codes in the census data.}
    \label{tab:suppdemographic}
    \end{table}
}

\newcommand{\suppbinggoogle}{
    \begin{figure}[h]
    \includegraphics[width=\linewidth]{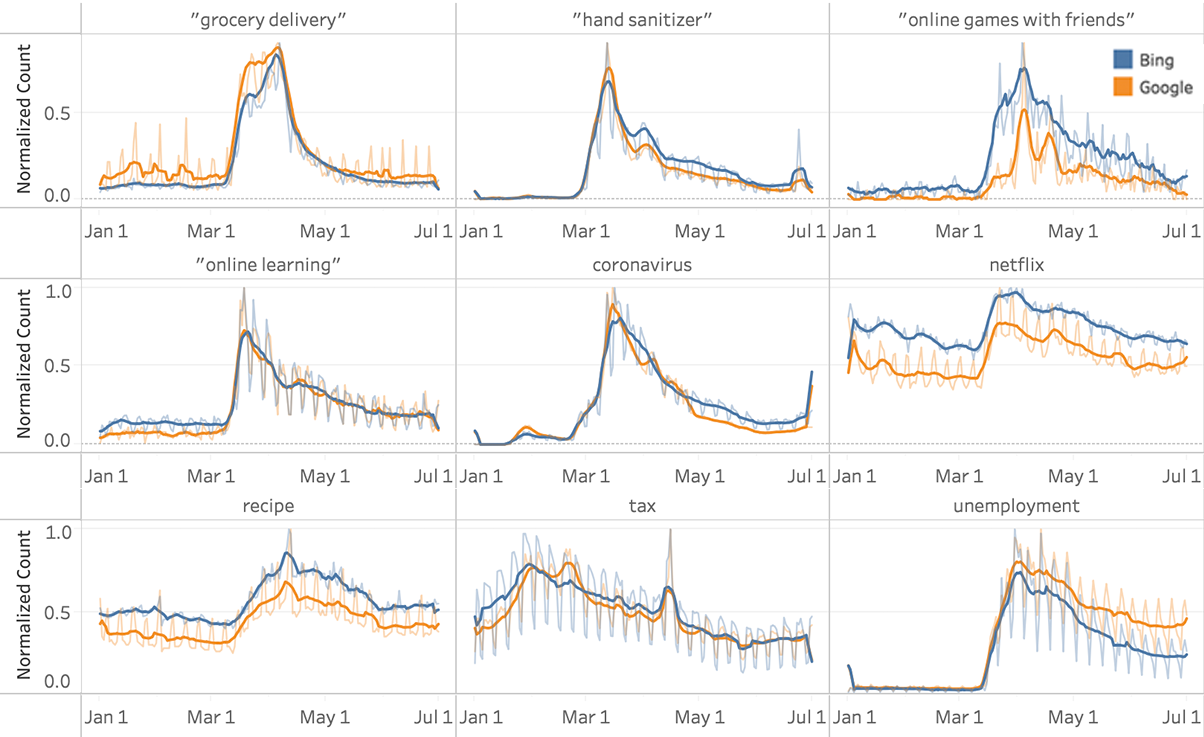}
    \vspace{0.1em}
    \captionsetup{width=\linewidth, labelfont=bf,textfont={bf},font=small,aboveskip=0pt, belowskip=0pt, singlelinecheck=false}
    \caption{Comparison of Bing and Google trends data on select keywords. The query counts are normalized where 1.0 denotes the maximum query count in each keyword trend. Bing trends (blue) closely follow Google trends (orange) with a median correlation of r=0.96.}
    \Description{Comparison of Bing and Google trends data on select keywords. The query counts are normalized where 1.0 denotes the maximum query count in each keyword trend. Bing trends (blue) closely follow Google trends (orange) with a median correlation of r=0.96.}
    \label{fig:binggoogle}
    \end{figure}
}

\newcommand{\suppsubneeds}{
\onecolumn
\begin{longtable}[h]{lllll}
    \toprule
    \textbf{Need Category} & \textbf{Need Id} & \textbf{Subcategory} & \textbf{Logic} & \textbf{Example} \\
    \midrule
    \multirow{32}{*}{SelfActualization} & SA1 & Audio/video related purchase & KD & home theater + e-commerce url click \\
     & SA2 & Books and reading related purchase & KD & comics + e-commerce url click \\
     & SA3 & Charity related queries & Q & charity; donations \\
     & SA4 & Cooking related purchase & KD & cookbooks + e-commerce url click \\
     & SA5 & Cooking related queries & Q & how to saute onions \\
     & SA6 & Cooking site visits & D & click on foodnetwork.com \\
     & SA7 & Crafts instruction queries & Q & how to embroider \\
     & SA8 & Crafts related purchase & KD & scrapbooking + e-commerce url click \\
     & SA9 & DIY site visits & D & click on instructables.com \\
     & SA10 & Home improvement purchase & KD & circular saw + e-commerce url click \\
     & SA11 & Home improvement related queries & Q & remodeling; wood working \\
     & SA12 & Home improvement site visits & D & click on homedepot.com \\
     & SA13 & Life goal related queries & Q & goals for living; personal goals \\
     & SA14 & Media related purchase & KD & dvd player + e-commerce url click \\
     & SA15 & Music instruction queries & Q & learn to play guitar; how to play a drum \\
     & SA16 & Musical instrument related purchase & KD & piano + e-commerce url click \\
     & SA17 & Outdoor related purchase & KD & hiking shoes + e-commerce url click \\
     & SA18 & Outdoor related queries & Q & best national parks \\
     & SA19 & Outdoor site visits & D & click on rei.com \\
     & SA20 & Parenting site visits & D & click on parents.com \\
     & SA21 & Pet related purchase & KD & ferret + e-commerce url click \\
     & SA22 & Pet related queries & Q & care for frogs \\
     & SA23 & Pet related site visits & D & click on petco.com \\
     & SA24 & Photography related purchase & KD & photography + e-commerce url click \\
     & SA25 & Photography related queries & Q & how to take pictures of a bird \\
     & SA26 & Sports related purchase & KD & elliptical trainer + e-commerce url click \\
     & SA27 & Streaming media site visits & D & click on netflix.com \\
     & SA28 & Technology related purchase & KD & ethernet + e-commerce url click \\
     & SA29 & Toys and gaming purchase & KD & xbox + e-commerce url click \\
     & SA30 & Toys and gaming site visits & D & click on twitch.tv \\
     & SA31 & Wedding related purchase & KD & wedding decorations + e-commerce url click \\
     & SA32 & Wedding site visits & D & click on theknot.com \\ \midrule
    \multirow{4}{*}{Cognitive} & C1 & Cognition related queries & Q & improve memory; cant pay attention \\
     & C2 & Educational degree related queries & Q & degree program; online diploma \\
     & C3 & Educational site visits & D & click on coursera.org \\
     & C4 & Online education queries & Q & learn remotely; lesson plans \\ \midrule
    \multirow{8}{*}{LoveBelonging} & LB1 & Dating related queries & Q & online dating; relationship advice \\
     & LB2 & Dating site visits & D & click on tinder.com \\
     & LB3 & Divorce related queries & Q & divorce lawyer; file for divorce \\
     & LB4 & Mental health experiential queries & Q & im alone; i feel depressed \\
     & LB5 & Mental health resource site visits & D & click on talkspace.com \\
     & LB6 & Online social activities queries & Q & online activities with family \\
     & LB7 & Social network site visits & D & click on facebook.com \\
     & LB8 & Social technology uses & KD & families + click on whatsapp.com \\ \midrule
    \multirow{15}{*}{Safety} & S1 & Bank related queries & Q & banks; banking \\
     & S2 & Bankruptcy related queries & Q & foreclosure;bankruptcy \\
     & S3 & COVID-19 protection purchase & KD & disposable masks; sanitizers \\
     & S4 & Domestic violence queries & Q & domestic assult; abusive relationship \\
     & S5 & Eviction related queries & Q & evicted; rent moratorium \\
     & S6 & Financial loan related queries & Q & borrower; mortgage \\
     & S7 & Firearm purchase & KD & glock holster + e-commerce url click \\
     & S8 & Housing related queries & Q & best neighborhoods; housing safety \\
     & S9 & Job search related queries & Q & job application; resume \\
     & S10 & Job search site visits & D & click on indeed.com \\
     & S11 & Rental related queries & Q & rental apartments; houses for rent \\
     & S12 & State unemployment site visits & D & click on www.michigan.gov/uia \\
     & S13 & Stimulus related queries & Q & relief fund; loan forgiveness \\
     & S14 & Tax related queries & Q & irs; tax \\
     & S15 & Unemployment related queries & Q & im unemployed; jobless benefits \\ \midrule
    \multirow{20}{*}{Physiological} & P1 & Apparel purchase & KD & athletic wear + e-commerce url click \\
     & P2 & Automobile related purchase & KD & motorcycle + e-commerce url click \\
     & P3 & Beverage purchase & KD & coffee + e-commerce url click \\
     & P4 & Cookware purchase & KD & cookie sheet + e-commerce url click \\
     & P5 & Food and beverage related queries & KD & applesauce; noodle soup \\
     & P6 & Food assistance related queries & Q & food stamps; snap program \\
     & P7 & Food delivery related queries & Q & grocery delivery; deliver meal \\
     & P8 & Food delivery site visits & D & click on instacart.com \\
     & P9 & Grocery related queries & Q & grocery; groceries \\
     & P10 & Grocery site visits & D & click on albertsons.com \\
     & P11 & Health condition related queries & Q & arthritis; diabetes \\
     & P12 & Health first aid purchase & KD & bandaid; wound closure \\
     & P13 & Health measurement equipment purchase & KD & oximeter + e-commerce site visits \\
     & P14 & Health symptom related queries & Q & i lost hearing; blurred vision \\
     & P15 & Household good purchase & KD & cleaning supplies + e-commerce url click \\
     & P16 & Insomnia related queries & Q & I cant sleep; help falling asleep \\
     & P17 & Prescription related queries & Q & medication interactions; pharmacies \\
     & P18 & Prescription site visits & D & click on rxlist.com \\
     & P19 & Sleep aid purchase & KD & sleep supplement; melatonin \\
     & P20 & Toilet paper purchase & KD & toilet paper; cottenelle \\
     \bottomrule
    \captionsetup{width=\linewidth, labelfont=bf,textfont={bf},font=small,aboveskip=0pt, belowskip=0pt, singlelinecheck=false} \\
  \caption{Human need categories and subcategories with examples. Under the Logic column, `KD' refers to matching both query string and clicked URL, `Q' refers to matching query string only, and `D' refers to matching clicked URL only. Need Id is provided to cross-reference with a full regular expression table provided in a separate file.}
  \label{tab:suppsubneeds} \\
\end{longtable}
\twocolumn
}
    
\newcommand{\suppdemcorr}{
    \begin{table}[h]
    \small
    \begin{tabular}{ll}
    \toprule
    \textbf{Demographic Variable} & \textbf{Corr Coeff} \\
    \midrule
    \% Housing Owned & -0.058*** \\
    \% Female & -0.029*** \\
    \% Race White & -0.024***\\
    Median Income & 0.021*** \\
    \% Has Internet & 0.019*** \\
    \% HS Grad or Higher & 0.018*** \\
    Median Age & -0.010 \\
    \% Below Poverty Level & -0.007 \\
    \bottomrule
    \multicolumn{2}{l}{*$p<0.05$, **$p<0.01$, ***$p<0.001$}
    \end{tabular}%
    \vspace{1pt}
    \captionsetup{width=\linewidth, labelfont=bf,textfont={bf},font=small,aboveskip=0pt, belowskip=0pt, singlelinecheck=false}
    \caption{Pearson correlation between Bing client rate and demographic variables for each ZIP code.}
    \label{tab:suppdemcorr}
    \end{table}
}

\newcommand{\suppbinggooglecorr}{
    \begin{table}[h]
    \small
    \begin{tabular}{lll}
    \toprule
    \textbf{Need} &\textbf{Keyword} & \textbf{Corr Coeff} \\
    \midrule
    SA & recipe & 0.960*** \\
    SA & netflix & 0.935*** \\
    SA & ``online games with friends'' & 0.896*** \\
    Cog & ``online learning'' & 0.973*** \\
    L\&B & ``online dating'' & 0.448*** \\
    Safe & unemployment & 0.977*** \\
    Safe & ``hand sanitizer'' & 0.966*** \\
    Safe & tax & 0.913*** \\
    Safe & gun & 0.764*** \\
    Phys & ``grocery delivery'' & 0.980*** \\
    Phys & coronavirus & 0.964*** \\
    Phys & ``food stamp'' & 0.963*** \\
    Phys & health & 0.888*** \\
    \bottomrule
    \multicolumn{2}{l}{*$p<0.05$, **$p<0.01$, ***$p<0.001$}
    \end{tabular}%
    \vspace{1pt}
    \captionsetup{width=\linewidth, labelfont=bf,textfont={bf},font=small,aboveskip=0pt, belowskip=0pt, singlelinecheck=false}
    \caption{Pearson correlation between Bing search trend and Google search trend for each keyword. Quotation marks indicate that the entire string is matched.}
    \label{tab:suppbinggooglecorr}
    \end{table}
}

\newcommand{\suppfigsubneeds}{
    \begin{figure}[h]
    \includegraphics[width=0.9\linewidth]{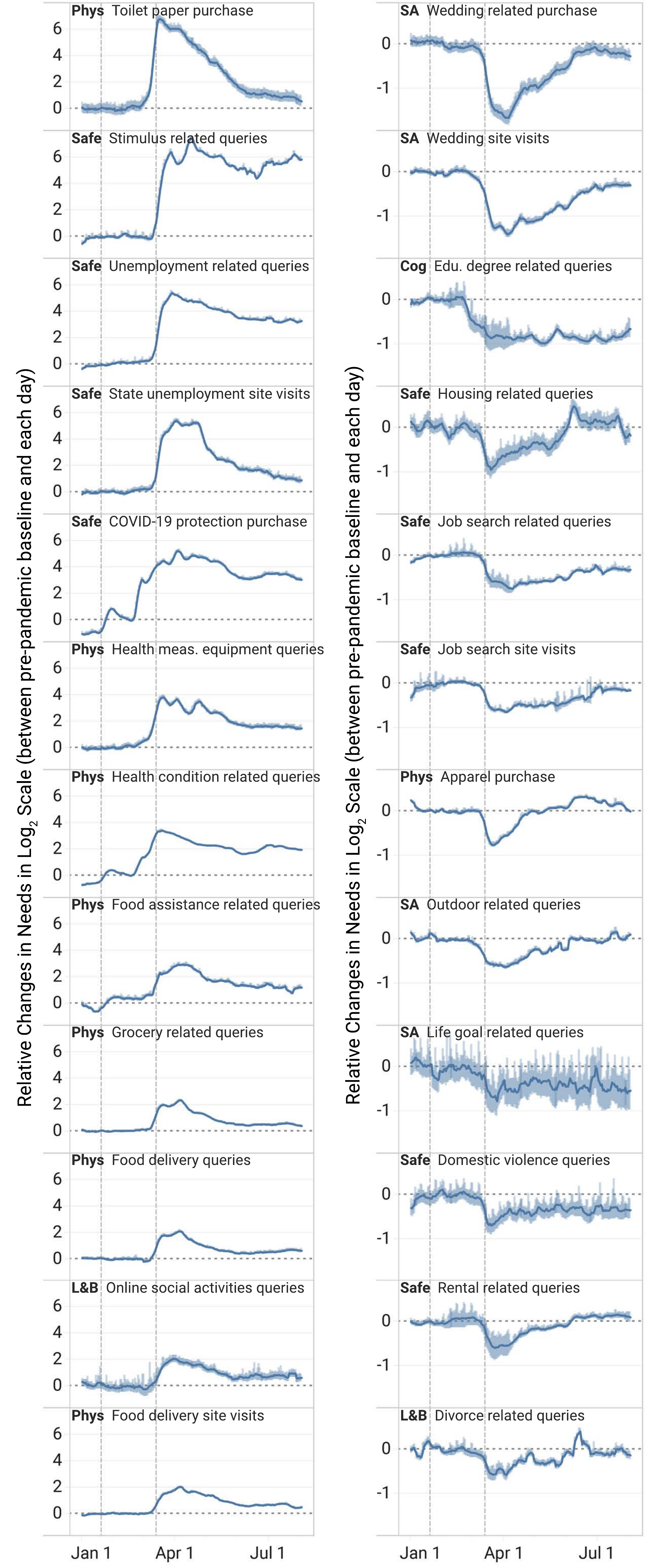}
    \vspace{0.5pt}
    \captionsetup{width=\linewidth, labelfont=bf,textfont={bf},font=small,aboveskip=0pt, belowskip=0pt, singlelinecheck=false}
    \caption{Daily relative changes in needs for top 12 need subcategories with the largest increase (left) and top 12 need subcategories with the largest decrease (right). Vertical bars denote the first reported US COVID case (Jan 20) and the US national emergency declaration (Mar 13).}
    \Description{Daily relative changes in needs for top 12 need subcategories with the largest increase (left) and top 12 need subcategories with the largest decrease (right). Vertical bars denote the first reported US COVID case (Jan 20) and the US national emergency declaration (Mar 13).}
    \label{fig:suppsubneeds}
    \vspace{-8pt}
    \end{figure}
}

\newcommand{\suppfigmturktask}{
    \begin{figure}[h]
    \includegraphics[width=1.0\linewidth]{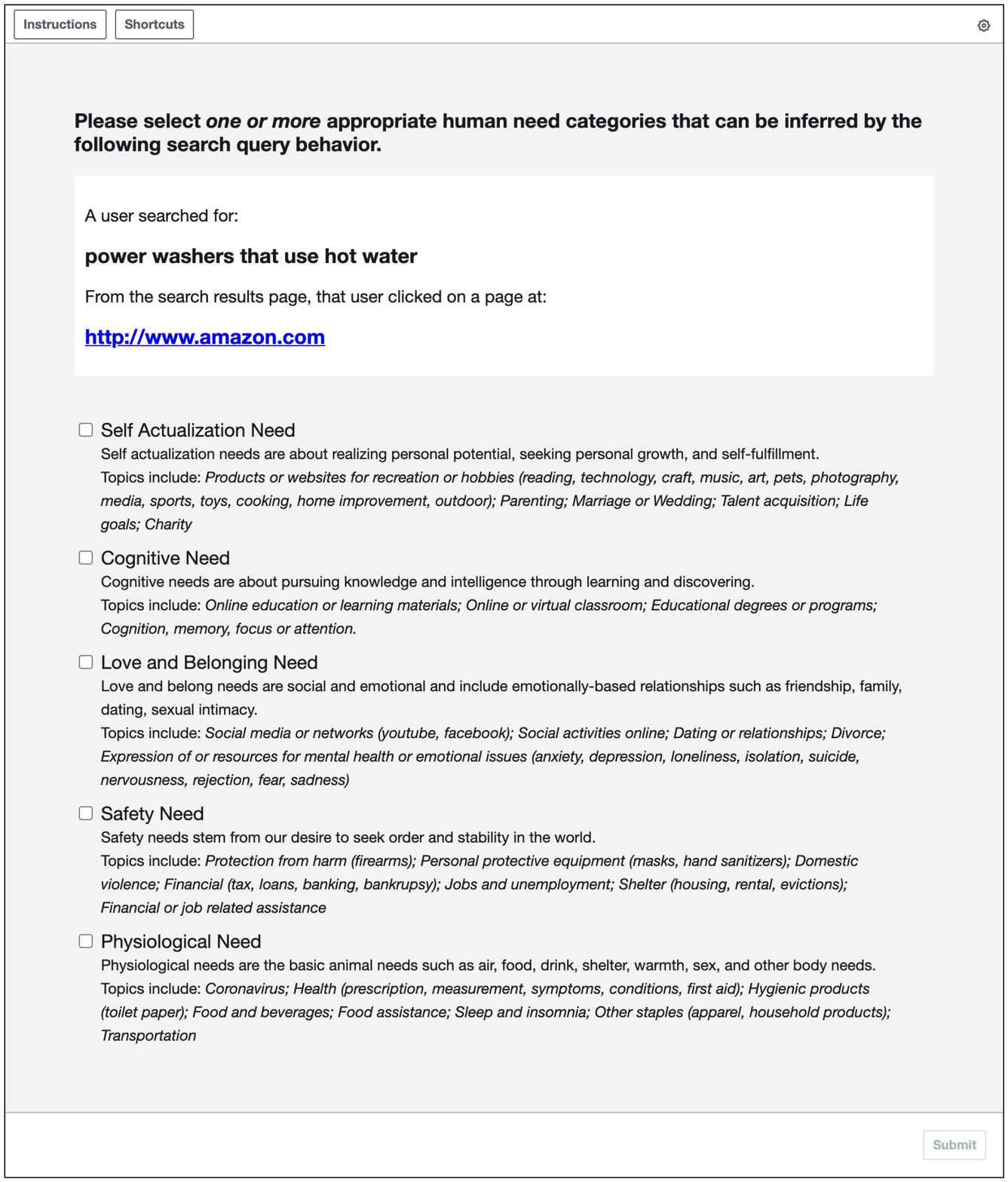}
    \vspace{0.5pt}
    \captionsetup{width=\linewidth, labelfont=bf,textfont={bf},font=small,aboveskip=0pt, belowskip=0pt, singlelinecheck=false}
    \caption{Example labeling task for Amazon's Mechanical Turk crowd worker. A query string and/or a clicked domain URL is displayed with multi-select options for five need categories. Each option includes a brief description and a set of example topics for the category.}
    \Description{Example labeling task for Amazon's Mechanical Turk crowd worker. A query string and/or a clicked domain URL is displayed with multi-select options for five need categories. Each option includes a brief description and a set of example topics for the category.}
    \label{fig:suppmturktask}
    \vspace{-8pt}
    \end{figure}
}

\begin{abstract} 
Most work to date on mitigating the COVID-19 pandemic is focused urgently on biomedicine and epidemiology. Yet, pandemic-related policy decisions cannot be made on health information alone. Decisions need to consider the broader impacts on people and their needs.  Quantifying human needs across the population is challenging as it requires high geo-temporal granularity, high coverage across the population, and appropriate adjustment for seasonal and other external effects.
Here, we propose a computational methodology, building on Maslow's hierarchy of needs, that can capture a \emph{holistic} view of relative changes in needs following the pandemic through a difference-in-differences approach that corrects for seasonality and volume variations. 
We apply this approach to characterize changes in human needs across physiological, socioeconomic, and psychological realms in the US, based on more than 35 billion search interactions spanning over 36,000 ZIP codes over a period of 14 months. 
The analyses reveal that the expression of basic human needs has increased exponentially while higher-level aspirations declined during the pandemic in comparison to the pre-pandemic period. 
In exploring the timing and variations in statewide policies, we find that the durations of shelter-in-place mandates have influenced social and emotional needs significantly. 
We demonstrate that potential barriers to addressing critical needs, such as support for unemployment and domestic violence, can be identified through web search interactions.
Our approach and results suggest that population-scale monitoring of shifts in human needs can inform policies and recovery efforts for current and anticipated needs.
\end{abstract}

\maketitle

\fancyhead{}

\enlargethispage{2\baselineskip}

\section{Introduction}
Many of the existing studies and datasets of the COVID-19 global pandemic focus on the biomedical and epidemiological aspect of the case and fatality rates, including efforts in detection, infection propagation, therapeutic intervention, and vaccine design, with a gaze fixed on the virus and illness that it causes.
Despite the direct focus on mitigating the spread and morbidity of infection~\cite{wellenius2020impacts}, pandemic-related policy decisions and investments cannot be made on health information alone.   
Challenges span societal (e.g., disparities~\cite{chowkwanyun2020racial}), economic (e.g., unemployment~\cite{baek2020unemployment,coibion2020labor}), and psychosocial (e.g., stress, anxiety~\cite{pfefferbaum2020mental}, loneliness~\cite{usher2020life}) realms. 
Recent work has called for identifying and understanding the multi-level system of humans needs and well-being for pandemic response and recovery strategies~\cite{ryan2020covid}.
Our goal is to better understand the influences of the pandemic and associated policy decisions on a multitude of human needs, where new insights about shifting needs can guide valuable refinements of policies and motivate the development of new interventions, programs, and investments.

Quantifying human needs across the population is important but challenging, as it requires innovative, ethical, privacy-preserving approaches with fine-grained and broad geo-temporal coverage.
A standard way to assess human needs is through survey-based measures~\cite{gallup1976human}, which can be costly and time-consuming to conduct at large scales.
Surveys are difficult to manage across time and geographies when the desire is to provide fine-grained analyses longitudinally and  to be able to understand and react in near real-time. 
Passively observing human behaviors is another approach, grounded in the fact that physical and psychosocial needs motivate human behaviors to express and fulfill those needs when they are unmet~\cite{maslow1943theory}.
For example, historical purchase behaviors are used to predict future consumer needs in market research~\cite{nielsen_2020}, but this approach is also limited to smaller-scale, consumer and commercial interests.
E-commerce platforms (e.g., Amazon marketplace) or specialized service providers (e.g., Talkspace, Coursera) may have access to large-scale, real-time analysis of customer behaviors, but they are focused tightly on  specific needs.
Publicly available social network data (e.g., Twitter), have been used to characterize needs~\cite{alharthi2017dataset,yang2013identifying}, but these studies examine a subset of needs from data that only portrays externalized behaviors.
Such fragmentation of data limits the capture of broader expressions and comparisons across a broad spectrum of human needs. 

We address these limitations in obtaining signals about human needs by observing the behaviors of people through their everyday interactions with a web search engine. 
Human behaviors, through which human needs are expressed or fulfilled, often involve seeking information or obtaining tangible support or material items, for which web search has been an integral component.
Thus, search logs provide a unique lens into human needs via providing signals about human behaviors in their natural state, at large scale, and on already routinely collected data.

\squeezesome{We propose a computational methodology built on constructs of human needs by Maslow~\cite{maslow1943theory,maslow1970new} and Max-Neef~\cite{max1992human}.
We characterize pandemic changes across a broad spectrum of \emph{fundamental human needs}, spanning five broad human needs categories--\emph{Self Actualization}, \emph{Cognitive}, \emph{Love and Belonging}, \emph{Safety}, and \emph{Physiological}--and 79 subcategories. 
We apply this framework to a dataset of 35+ billion search interactions across 36,000+ ZIP codes in the United States and over 14 months (7 months in 2019 and 7 months in 2020) to map search query strings and click interactions to human needs, resulting in over three billion expressions of human needs.
We demonstrate how this approach enables the examination of shifts in fundamental human needs based on disruptions induced by the pandemic.}

Our contributions include the following:
\begin{itemize}
    \item \squeeze{We propose a novel computational framework for characterizing a \emph{holistic} view of human behaviors, intents, and unmet needs based on web search logs and human motivational theories (Sec.~\ref{sec:humanneedsframework})}.
    \item \squeeze{We leverage a difference-in-differences approach~\cite{lechner2011estimation}} to quantify the impact of the pandemic and its associated policies on the relative changes in needs while controlling for seasonality and external factors (Sec.~\ref{sec:quantneed}).
    \item We present the first population-scale analysis across a \emph{holistic} set of human needs during the COVID-19 pandemic in the United States through the use of web search logs (35+ billion search interactions for 14 months on 86\% of US ZIP codes; Sections~\ref{sec:tempchange}-\ref{sec:consequences}).
    \item We find that search interactions in pursuit of basic human needs (i.e., \emph{Physiological}, \emph{Safety}) have  increased exponentially during the pandemic while several higher-level aspirations (i.e., \emph{Self Actualization}, \emph{Cognitive}) have declined (Sections~\ref{sec:tempchange},~\ref{sec:subneeds}).
    \item We observe geographical differences in how differing statewide shelter-in-place policies are associated with short-term and long-term changes in social and emotional needs (Sec.~\ref{sec:geo}).
    \item We demonstrate that potential barriers to accessing critical resources, in support of people facing unemployment or domestic violence, can be identified through search interactions combined with external data sources (Sec.~\ref{sec:consequences}).
\end{itemize}

Our work suggests that signals from web search logs can be used to characterize and to monitor over time human needs at a population scale.
Our findings also emphasize the importance of tracking broad sets of human needs in combination with other reported measures to identify gaps in our current understanding of challenges and support, to measure the impact of policy changes, and to design policies and programs that can meet the needs.

\section{Related Work}

\enlargethispage{2\baselineskip}

\xhdr{Quantifying Human Needs}\label{back:needs}
\squeeze{Theories about basic human needs have been discussed for close to a century~\cite{pittman2007basic}. In particular, Maslow's hierarchy of needs~\cite{maslow1943theory,maslow1970new} has been applied in numerous domains~\cite{yang2013identifying,van2015understanding,cerbara2020nation, moss2010introduction}, despite criticisms of the validity of the theory~\cite{wahba1976maslow}. 
These theories are aimed at providing a holistic understanding of human needs. The characterizations of broad spectrum of needs from these models are increasingly relevant during the pandemic~\cite{ryan2020covid}.}
\squeeze{Most studies use survey-based methods to measure human needs~\cite{cerbara2020nation,tay2011needs,mathes1978empirical,mitchell1976measurement}.}
Others have applied human needs theories in computational social science~\cite{alharthi2017dataset,yang2013identifying,long2020needfull}, but they focus deeply on specific topics (e.g., consumer behavior, well-being) and leverage publicly available social network data.
In contrast, we introduce a computational methodology to extract a full spectrum of human needs at population scales, which is critical for aligning policies with societal needs that they are intended to support. 

\xhdr{Web Search Logs for Human Needs Analysis}
In addition, we harness search interaction data which captures more natural observations of human needs~\cite{dumais2014understanding}.
Web search logs have been used to understand human behaviors across many different domains~\cite{althoff2018psychomotor,ginsberg2009detecting,west2013cookies,weber2012mining}, time~\cite{althoff2016influence,althoff2017harnessing,paul2016search,fourney2015exploring}, location~\cite{west2013here,sadilek2018machine}, and to predict the now and the future~\cite{d2010google,choi2012predicting}, but prior studies typically focus on a single aspect of human well-being. 
Google Trends APIs have helped to stimulate a prolific range of research in the context of the COVID-19 pandemic for physical~\cite{lin2020google}, psychological~\cite{tubadji2020narrative}, and socioeconomic~\cite{gupta2020effects,abay2020winners} well-being that operate on highly normalized and aggregated data. 
In distinction, our work leverages fine-grained geospatial comparisons across 79 need subcategories and a difference-in-differences methodology, which allows for controlling confounders and understanding national coverage with flexible geo-temporal normalization and aggregation. We also improve detection of needs by leveraging click interactions.

\enlargethispage{\baselineskip}
\section{Dataset and Validity}\label{sec:data}

\process

\subsection{Dataset, Privacy, and Ethics}
We collected a dataset containing a random sample of deidentified search interactions from the first seven months of the years 2019 and 2020 obtained from Microsoft's Bing search engine.
For each search interaction, we collected the search query strings, all subsequent clicks from the search results page (e.g., clicked URL), and time and ZIP code location of the search interaction.
The resulting dataset contains 35+ billion search interactions and represents the web search traffic of over 86\% (36,667/42,632) of US ZIP codes associated with at least 100 queries per month so as to preserve anonymity (Table~\ref{tab:dataset}). 
All data was deidentified, aggregated to ZIP code levels or higher, and stored in a way to preserve the privacy of the users and in accordance to Bing's Privacy Policy.
\irb
\tabdataset


\subsection{Validation of Data}
Considering potential threats to validity, we examined the dataset from three perspectives: coverage of the population, analysis of selection bias, and reliability of the trends.

\xhdr{Analyzing National Representation}
To understand how much of the US population is represented by the collected data, we obtained demographics data from the Census Reporter API~\cite{census_reporter}. 
The demographics of the ZIP codes in our dataset closely matched the US population demographics (\onlineappendix{A.1}). 
Although query volumes are not uniformly distributed across these ZIP codes, the vast majority of the ZIP codes are included in our dataset. 
We leverage location information for our analysis when geographical differentiation is necessary.

\xhdr{Analyzing Selection Bias} 
To understand potential biases in socioeconomic circumstances that would influence the usage of the Bing search engine, we leveraged deidentified client id as a proxy for a unique user to estimate the `client rate', or how much of the population in a ZIP code is using the Bing search engine.
We examined the correlation between the client rate and various demographic factors.
While the factors describe some of the variance, none were correlated more strongly than $r$=--0.058 (\% Housing Owned), suggesting that the dataset is not strongly biased towards any single demographic (\onlineappendix{A.2}). 

\xhdr{Reliability of Search Interaction Trends}
Many Americans use other search engines such as Google. 
Therefore, we compared search trends for Bing with data available via the Google Trends API for the same time period and for specific keywords in each need category.
We  found that the search trends are highly similar, with a median Pearson correlation of 0.96 (min=0.45, max=0.98, all $p$<0.001). 
\squeeze{This implies that our findings are not simply an artifact of using one search engine over another (\onlineappendix{A.3}).} 

\section{Human Needs Framework}\label{sec:humanneedsframework}

\subsection{Human Needs Categories}\label{method:categories}
We draw inspiration from Maslow's hierarchy of fundamental human needs~\cite{maslow1943theory,maslow1970new} to tag each search interaction with one or more of five broad categories of needs. We represent human needs as a ladder to convey that a person may have multiple needs (Fig.~\ref{fig:process}a). \emph{Safety} and \emph{Physiological} are considered as `basic' needs. \emph{Love and Belonging} is often considered to be `psychological' needs, and \emph{Cognitive} and \emph{Self Actualization} are considered as `growth' needs, defined in more detail below:
\enlargethispage{2\baselineskip}

\textbf{Self Actualization} needs are about realizing personal potential, seeking personal growth, and self-fulfillment. Topics include: hobbies; parenting; wedding; talent acquisition; goals; charity.

\textbf{Cognitive} needs are about pursuing knowledge and intelligence through learning and discovering. Topics include: online education, learning materials; educational degrees; cognition; memory; focus.

\textbf{Love and Belonging} are social and emotional needs and include emotionally-based relationships such as friendship, family, dating, sexual intimacy. Topics include: mental health or emotions; social network or activities; relationships, dating, divorce or breakup.

\textbf{Safety} needs stem from our desire to seek order, stability, and protection from elements in the world. Topics include: personal protection; finances; banking; job search; unemployment; housing.

\textbf{Physiological} needs are the basic animal needs such as air, food, drink, shelter, warmth, sex, and other body needs. Topics include: health; food and groceries; basic staples; sleep; transportation.

\squeeze{To understand the nuances of the needs, we further subdivided the five main categories into 79 subcategories. 
Several researchers independently developed subcategories which were combined and resolved collaboratively through consensus meetings.
\onlineappendix{A.4} describes this process in detail with the full taxonomy of our need categories, example queries and/or clicked page URLs.} 

\subsection{Human Needs Detection}\label{method:processing}

\enlargethispage{\baselineskip}

A search interaction can be an observation of the underlying human need in two ways: (1) an expression of a potential satisfier (physical or information) for that need, or (2) a direct expression of the deficiency or satisfaction of that need. 
For example, a search query for `bandages' with a subsequent click on `amazon.com' could indicate a purchase intent that satisfies a \emph{Physiological} need. 
We require the additional click into one of many e-commerce domains to solidify that this interaction is a purchase intent.
Information search about `online games with friends' could satisfy a \emph{Love and Belonging} need.
A need (satisfaction or deficiency thereof) could be directly expressed in experiential statements such as `I feel depressed' (\emph{Love and Belonging}). 

We match each search interaction to a corresponding need subcategory through simple detectors based on regular expressions and basic propositional logic. 
Each need subcategory could have multiple regular expressions applied to either the query string, the clicked URL, or both, depending on the complexity of the expression and the need subcategory. 
We arrived at these regular expressions based on our data through several collaborative consensus meetings until we were satisfied with precision and recall (\onlineappendix{A.4}). 
Overall, 9.1\% of our query samples matched at least one of the need categories.
Each search interaction can satisfy multiple human needs~\cite{max1992human}, so we allowed each search interaction to be tagged by multiple need categories
(only 0.32\% have multiple tags).
We then aggregated matched search interactions across need categories and subcategories, time (e.g., day, week), and geography (e.g., ZIP code, county, state). 
Fig.~\ref{fig:process} illustrates these steps in detecting and processing of human needs. 

Throughout the paper, we report needs as \emph{expressed} through search interactions and not actual underlying human needs. 
Some human needs are expressed well by search interactions while other needs are more appropriately expressed through other digital and/or non-digital means. 
When analyzing \emph{changes} in such search interactions, careful consideration is required to differentiate actual shifts in the underlying human needs (e.g., health-related needs have increased due to COVID-19) from the equally meaningful changes observed in logs based upon shifts from offline to online behaviors (e.g., online grocery purchases due to store closures), and we present our results in both possible contexts.

\subsection{Framework Validation}\label{sec:validation}
Our goal is to detect with high precision a large number of needs across a broad set of categories. 

\enlargethispage{2\baselineskip}

\xhdr{Precision}
We sampled ~1.2 million search interactions that matched at least one need category as a candidate set.
From this sample, we randomly chose 100 unique tuples of search query string and clicked URL (e.g., `15 lb dumbbells' and click on `walmart.com') for each of the five high-level need categories, for a total of 500 unique tuples representing 1,530 search interactions in our evaluation set.
We selected unique tuples to avoid duplication in labeling, but we mapped the labels back to the original 1,530 search interactions to compute precision on the distribution of the source evaluation set. 
We then collected human labels for each tuple via Amazon's Mechanical Turk, where each tuple could be tagged with none, one, or more of the five needs categories. 
All labels and predictions have Boolean values with no ranking among needs categories.

Upon inspecting the label quality, we found common systematic label errors such as labeling `recipe'-related queries as \emph{Cognitive} needs, or `divorce'-related queries or visits to specific government unemployment sites as \emph{Physiological} needs, where the workers mislabeled the queries according to the definitions we specified in the task detail. 
Other errors were due to inherent ambiguities in search. For example, `rent in florida coronavirus' is tagged as \emph{Physiological} for `coronavirus' but not as \emph{Safety} for `rent' because our high precision detector requires more qualified keywords such as `apartment rent.' 
Although the worker tagged this as \emph{Safety} (i.e., rent for shelter), the use of the word `rent' here may not be shelter-related. 
We took a conservative approach of only correcting definitive label errors and not ambiguous errors, and our evaluation set achieved a precision of $97.2\%$, using the example-based precision metric defined in~\cite{zhang2013review} for multi-label classification. 

\xhdr{Recall} 
Although it is infeasible to ensure a perfect recall across a massive dataset, it is important that we capture a significant number of needs expressions.
We find that 9.1\% of our search interactions match at least one need category.
While this recall is significant and led to more than 3.2 billion detections of needs expressions, 
we note that high recall is not necessary for an unbiased analysis approach, because we conduct a fair comparison among the outputs of the same detectors across pre-pandemic and pandemic periods.
We did not find that the exact expressions of needs varied drastically within our dataset that would indicate any temporal bias.
We also investigated whether our need expressions were dominated by a few categories. Clicks to YouTube or Facebook dominated, but still only represented 1\% of our dataset. 
We categorized visits to these sites based on their primary functions (i.e., Facebook for social networking and YouTube for media consumption). 
We found that our results were robust, whether or not we included these high-traffic sites in our dataset.


\subsection{Quantifying Changes in Human Needs}\label{sec:quantneed}
Our goal is to quantify the change in human needs during the pandemic relative to the pre-pandemic period.
This can be challenging due to potential confounding effects of yearly seasonal variations, weekly seasonal variations, and variations in query volume over time. 
Conceptually, we control for yearly seasonal effects through comparisons with the previous year, for weekly seasonal effects by matching the day of the week between both years (i.e., \mbox{Mon Jan 6, 2020} is aligned to \mbox{Mon Jan 7, 2019}), and by considering relative proportions of the query volume represented by each need over time.
Formally, we follow a difference-in-differences methodology~\cite{lechner2011estimation,dimick2014methods}, commonly used in economics, to account for confounding effects between comparison groups. 
Finally, our adjusted effect size is the logarithm of the ratio between two groups.
This is effectively the difference-in-differences approach applied to the logarithmic effect sizes and has the advantage of the effect sizes having symmetric properties (i.e., $\Delta(t_1; t_2)$=$-\Delta(t_2;t_1)$)~\cite{graff2014expressing, cole2017statistics}. 
This step allows for appropriate comparison of effect sizes across both increases and decreases in need.
Our estimate of the \emph{relative change in human need} $C$ between two time periods is defined as
\begin{displaymath}
C(t_1;t_2, n) = \log_2\left(\frac{E(t_2^{2020}, n)}{E(t_1^{2020}, n)}\right) - \log_2\left(\frac{E(t_2^{2019}, n)}{E(t_1^{2019}, n)}\right)
\end{displaymath}
where $E(t_{2}^{2020}, n)$ is the expression of need $n$ at some time $t_2^{2020}$ in 2020 (i.e., after the pandemic declaration) and $E(t_{1}^{2020}, n)$ is the expression of need at $t_{1}^{2020}$ (i.e., before the pandemic).

\squeeze{Across all following analyses, we choose the mean daily expression of needs between Jan 6 to Feb 23, 2020 as the `pre-pandemic baseline', referred to throughout the paper, and dates on or after Mar 16 as the `pandemic period' because individual states declared a state of emergency at different times (Feb 29 to Mar 15).} 
We then compute the 95\% confidence interval on this multiplicative effect size by using bootstrap resampling with replacement (N=500).
We report mean estimates and p-values throughout the text and 95\% confidence intervals in all figures and tables where applicable. 
All time series figures (Figures~\ref{fig:annotatedtrend}, \ref{fig:subneeds}, \ref{fig:unemployment}B) indicate the moving average of the daily relative changes, computed from 3 days before to 3 days after.

\section{Temporal Changes in Human Needs}\label{sec:tempchange}

\enlargethispage{2\baselineskip}
\annotateedtrend

We first consider how human needs change over time across the US in the context of major events surrounding the pandemic. 
We compute the daily relative change in the expressed needs in comparison to the pre-pandemic baseline, as described in Sec.~\ref{sec:quantneed}, for the duration of our entire dataset, giving us per-day relative changes in all need categories and subcategories.
For each inflection point and major national event, we examine need subcategories with the highest relative changes to understand which contribute the most to the overall need.

\xhdr{Elevated Needs and Contributing Subcategories}
Fig.~\ref{fig:annotatedtrend} illustrates daily relative changes of needs on a log scale, where zero indicates no change.
Overall, we see that all need categories were at elevated rates during March through May relative to the earlier months. 
A few of the local inflection points correspond to US national events, such as the declaration of national emergency on Mar 13 or the first stimulus checks being deposited on Apr 11. 

\emph{Physiological} needs start to increase first around February (Fig.~\ref{fig:annotatedtrend}A), dominated by \emph{health condition related queries} ($C$=$1.46$ on Feb 29) and subsequently by \emph{toilet paper purchase} and \emph{health measurement equipment purchase} ($C$=$1.14$, $0.76$ on Mar 6 respectively; Fig.~\ref{fig:annotatedtrend}B).
Around \mbox{Mar 16}, \emph{Physiological} needs peak at over 3.8 times the baseline ($2^{1.91}$; Fig.~\ref{fig:annotatedtrend}C). 
Following national emergency declaration (Mar 13) and mandated lock downs (first on Mar 21), we see a sharp increase in \emph{Cognitive} needs (Fig.~\ref{fig:annotatedtrend}D), dominated by \emph{educational site visits} and \emph{online education queries} ($C$=$1.97$, $1.42$ on Mar 23).
\emph{Self Actualization} needs peak around Apr 11 (Fig.~\ref{fig:annotatedtrend}E), dominated by \emph{cooking site visits} and \emph{cooking related queries} ($C$=$1.73$, $1.15$), and \emph{online social activities queries} and \emph{social technology uses} dominate \emph{Love and Belonging} needs ($C$=1.77, 1.72 on Apr 11).
A sharp spike of \emph{Safety} needs can be seen shortly after the first stimulus checks were deposited: \emph{stimulus related queries}, \emph{state unemployment site visits}, \emph{COVID-19 protection purchase} dominate \emph{Safety} needs \mbox{($C$=8.17, 5.49, 5.12 on Apr 18; Fig.~\ref{fig:annotatedtrend}F).}
While other needs start to trend downwards or stabilize throughout much of May-July, \emph{Physiological} needs increase for a second time with additional interests in health conditions, followed by \emph{Safety} needs with queries related to economic stimulus and loans (Fig.~\ref{fig:annotatedtrend}G), which aligns with the rise of COVID-19 cases in the US around Jun 6.

\enlargethispage{\baselineskip}

\xhdr{Shifting of Needs}
Based on the severe health impacts of the COVID-19 pandemic, we expected to see and confirmed that \emph{Physiological} needs dominate throughout our dataset, as COVID-19 is still a major US public health issue at the time of writing.
At a glance, we see two instances of the surge in \emph{Physiological} needs followed by a subsequent increase in \emph{Safety} needs. 
As \emph{Physiological} concerns rise, public health responses (e.g., business closures or restrictions) could potentially induce instabilities in \emph{Safety} needs, and this observation needs to be further investigated. 
We see basic needs expressed before other needs consistent with the hypothesis by Maslow~\cite{maslow1943theory} and observations by others~\cite{tay2011needs}. 
We also expected to see a decrease in the expression of growth needs (\emph{Self Actualization} and \emph{Cognitive}) as people's attention shifts toward basic needs. 
However, both \emph{Cognitive} and \emph{Self Actualization} needs increased overall, with the increase in \emph{Cognitive} needs being more temporary and \emph{Self Actualization} being more sustained. 
Despite health and economic concerns, interests in recreational activities or hobbies (e.g., cooking, gaming) contribute to this steady 23\% (=$2^{0.3}$--1) increase in \emph{Self Actualization} needs.
Further research into the temporary nature of \emph{Cognitive} needs and the long-term impact of such sustained interest in \emph{Self Actualization} is necessary.

\squeeze{We see that the peak in \emph{Physiological} needs occurs around four weeks before the peak in \emph{Safety} needs (Fig.~\ref{fig:annotatedtrend}C, F), while the second set of peaks are a few days apart (Fig.~\ref{fig:annotatedtrend}H). 
This could be an indication of phenomena like resilience or endurance from economics and disaster management that requires further examination~\cite{martin2015notion,fletcher2013psychological}.}
\section{Significant Changes in Human Needs}\label{sec:subneeds}

\enlargethispage{\baselineskip}

Next, we examine individual need subcategories that present the largest increase or decrease in search expressions, possibly due to the pandemic impact.
To explore these two ends of the spectrum, we compute the mean relative changes in needs during the initial four weeks of the pandemic period (Mar 16 to Apr 12) compared to the pre-pandemic baseline, as described in Sec.~\ref{sec:quantneed}. 
We then examined the top 12 need subcategories with the largest increase or decrease in the relative change.

\tabtopchange
\tabbottomchange

\xhdr{Heightened \emph{Physiological} and \emph{Safety} Needs}
\squeeze{Table~\ref{tab:topchange} shows that 11 need subcategories with the most increase fall under \emph{Physiological} and \emph{Safety} needs as seen from the temporal trends in Sec.~\ref{sec:tempchange}, and one (\emph{online social activities queries}) belongs to the \emph{Love and Belonging} need category.
\emph{Toilet paper purchase} reached a maximum increase of 127 times the pre-pandemic baseline (Mar 16).
Recall that these are not just queries containing `toilet paper', but purchase intents as indicated by subsequent clicks to e-commerce sites (Sec.~\ref{method:processing}).
Such a high level of interest in toilet papers is commonly attributed to panic buying due to the supply scarcity~\cite{hamilton2020scarcity} and media coverage~\cite{garfin2020novel}. 
\emph{Stimulus related queries}, including general terms like `loan forgiveness,' reached an even higher maximum increase of 286 times the baseline
(Apr 18) and is sustained at that high level through July, reflecting the magnitude of the pandemic's impact on the US economy.}

When we examine the daily trends, we see that \emph{COVID-19 protection purchase} exhibits a small peak ($C$=1.0 on Jan 29) after the first reported COVID-19 case (Jan 20), and the needs quickly escalate from Feb 20 (Fig.~\ref{fig:subneeds}B), at least three weeks earlier than other needs (Fig.~\ref{fig:subneeds}A,C,D) that do not escalate until the national emergency declaration. 
\squeeze{A subsequent peak on Apr 3 ($C$=5.6) coincides with CDC's updated recommendation on cloth-based mask use. }
\squeeze{We also find that indicators of social-economic instabilities such as \emph{unemployment site visits} and \emph{food assistance related queries} still have not returned to their baseline levels (Fig.~\ref{fig:subneeds}A,C), arguably because the pandemic is still in effect. 
\emph{Online social activities queries} follow a similar pattern (Fig.~\ref{fig:subneeds}D), reflecting the need to satisfy lock-down induced social isolation through online services, but it also raises the question of potential permanent shifts in ways of satisfying social needs.
Our findings revealed that only a few of these needs have returned to the pre-pandemic baselines while many of them are sustained at elevated rates.}

\subneeds

\xhdr{Shifts Away from Positive Outlooks}

\enlargethispage{\baselineskip}

Table~\ref{tab:bottomchange} shows the most decrease in the expression of several \emph{Self Actualization} and one \emph{Cognitive} need subcategories. 
Specifically, indications of \emph{Self Actualization} needs for partnership have declined by more than 64\% (=$2^{-1.49}$--1) of their baseline throughout the typical US wedding season around Spring and early Summer, which is expected given restrictions on large gatherings. 
In addition, needs that are typically associated with growth, positive outlook, or new opportunities have taken a large toll.
In \emph{Self Actualization}, queries about life goals also see a large decline.
In other need categories, needs expressed by \emph{educational degree related queries}, \emph{job search related queries}, \emph{job search site visits}, or \emph{housing related queries} have declined by over 34\% (=$2^{-0.61}$--1) of their baseline.

Upon inspection of daily trends, expressions of forward-looking needs have decreased and remain below the pre-pandemic baseline (Fig.~\ref{fig:subneeds}E,F,G). 
The sustained decline in \emph{job search related queries} (Fig.~\ref{fig:subneeds}G) juxtaposed with near 30 times increase in unemployment needs (Fig.~\ref{fig:subneeds}A) is a troubling evidence of the declining labor force as seen in other studies~\cite{coibion2020labor}.
Indications of growth interests in educational degrees or life goals have not recovered (Fig.~\ref{fig:subneeds}F). 
These results combined with heightened \emph{Physiological} needs suggest a shift of focus away from individual growth.  
\squeeze{\emph{Divorce related queries} exhibited a maximum of 47\% (=$2^{-0.93}$; Fig.~\ref{fig:subneeds}H) decline, possibly reflecting the challenges that families face in proceeding with divorce during the lock downs~\cite{lebow2020challenges}.
Therefore, underlying mechanisms for these shifts, long-term impact of the lack of growth needs, and support for relationships should be further studied.
See \onlineappendix{A.5} for corresponding figures on all 24 subcategories.} 

\section{Geographical Differences in Needs}\label{sec:geo}

\enlargethispage{\baselineskip}

We shift our attention to how the pandemic and its related policies \emph{differentially} influence local \emph{subpopulations}. 
We use a set of statewide policies\footnote{Although our dataset allows ZIP code level analysis, a comprehensive list of local policies across the US are difficult to obtain at ZIP code, city, or even county levels.} readily available through the COVID-19 US State Policy Database~\cite{raifman2020covid}. 
We examine the impact of the shelter-in-place mandate (its duration and effective date) on social-emotional and relationship needs (i.e., eight \emph{Love and Belonging} and two wedding-related \emph{Self Actualization} subcategories). 
We hypothesized that social isolation induced by longer shelter-in-place mandates generates more expressions of social-emotional needs.

\squeeze{We used the date on which the shelter-in-place mandate was enacted and relaxed or lifted for 38 states to derive the mandate duration\footnote{Only 38 states had the start and end dates in the dataset as of this writing.}. 
We compute two relative changes in needs for each state.
First, to understand the \emph{short-term} impact of the {shelter-in-place mandate}, we compute the relative change in needs for {\emph{two weeks after}} each state's shelter-in-place mandate compared to the {\emph{one week before}} the mandates\footnote{We chose one week before the mandate because it is the maximum number of full weeks after the declaration of national emergency and before the earliest start date (Mar 21) for any shelter-in-place mandate.}. 
Second, to understand the {\emph{long-term}} impact of the mandate, we compute the relative change in needs between the \emph{pre-pandemic baseline} and the \emph{last four weeks} of our dataset (Jul 6 to Aug 2), which is at least two weeks after the last state lifted its mandate (Jun 19).
To quantify the potential impact of these mandates, we ran a Pearson correlation analysis (1) between the start date of the shelter-in-place mandates (i.e., ISO day number) and the short-term relative changes in needs, and (2) between the duration of the mandates and the long-term relative changes in needs.}

\xhdr{Early Adjustments to Social Needs}
The start date of shelter-in-place mandates ranged between Mar 21 and Apr 7.
We find that the relative changes in \emph{online social activities queries} (\mbox{$r$=$-0.53$}, $p$<0.001), \emph{wedding site visits} (\mbox{($r$=0.48}, $p$=0.002), \emph{wedding related purchase} ($r$=0.43, $p$=0.006), and \emph{mental health resource site visits} ($r$=0.47, $p$=0.003) needs were most correlated with the start date.
People from states that have earlier shelter-in-place mandates expressed significantly reduced interests in weddings and mental health site visits (Fig.~\ref{fig:stateshelter}A) and significantly more need for online social activities (Fig.~\ref{fig:stateshelter}B).
For example, in the two weeks after the mandate, people in New Jersey (mandate on Mar 21) sought online mental health resources 25.7\% less than the week before the mandate, while people in South Carolina (mandate on Apr 7) sought those resources 13.4\% more than the week before.

\stateshelter

\xhdr{Long-term Social Impact}
\squeezetiny{When we examine the long-term changes in expressed needs, we find that shifts in \emph{negative mental health experiential queries} ($r$=0.42, $p$=$0.010$, Fig.~\ref{fig:stateshelter}C), \emph{wedding site visits} ($r$=$-0.37$, $p$=0.022, Fig.~\ref{fig:stateshelter}D), and \emph{wedding related purchase} ($r$=$-0.35$, $p$=0.033) were most correlated with the duration of sheltering and closures.
The duration of the shelter-in-place mandates were highly correlated with the start date ($r$=$-0.62$, $p$<0.001), indicating that people faced with earlier mandates are also impacted by longer mandates. 
As we discovered above, these people likely suppressed their needs for weddings or mental health support early during the pandemic.
At the same time, as wedding needs slowly recover to the pre-pandemic baseline (Fig.~\ref{fig:subneeds}E), those impacted by longer mandates are even slower in their recovery of wedding needs and are more likely to express negative mental health issues.
For example, people in Mississippi (24 days of shelter-in-place) expressed 33.2\% \emph{less} negative mental health experiences than before the pandemic while people in Oregon (88 days) expressed 27.2\% \emph{more} negative mental health experiences than before (Fig.~\ref{fig:stateshelter}C). 
Others attribute this increase in negative emotions during the pandemic to the shift to basic needs~\cite{cerbara2020nation}. 
Per-state analysis of this shift along with differential prevalence of COVID-19 may be necessary to understand the mechanisms for this increase in negative emotions and to provide appropriate social-emotional support.}

\enlargethispage{\baselineskip}
\section{Gaps between Expressed and Reported Needs}\label{sec:consequences}
\squeeze{As we have demonstrated so far, there are many needs that are well expressed by web search interactions such as purchasing goods online, looking for health information online, or accessing economic assistance through government websites.
Our final analysis examines a gap between how web search facilitates the expression of these needs and the reported fulfillment of these needs to discover potential barriers to accessing critical resources.
To demonstrate how our approach allows for exploring these barriers, we focus on two need subcategories: unemployment and domestic violence.}

\enlargethispage{2\baselineskip}

\unemployment
\xhdr{\mbox{Expressed Unemployment Needs vs. Reported Claims}}
We obtained weekly, seasonally adjusted initial unemployment claims for 2020 from the US Department of Labor\footnote{\url{https://oui.doleta.gov/unemploy/claims.asp}} and computed the relative changes in unemployment claims using the same approach as in Sec.~\ref{sec:quantneed}.
We compare the change in reported unemployment claims with the change in expression of unemployment needs in search and find that these two changes are closely aligned (Pearson $r$=0.996, $p$<0.001) (Fig.~\ref{fig:unemployment}A).
Despite policies that extend unemployment eligibility for up to 26 weeks during the pandemic, we see that the expression of unemployment needs remain at 25\% higher than the reported claims since April.
This discrepancy corroborates with known issues with the unemployment benefits: many people were confused about the benefits (e.g., job search requirement) and were being denied, having to search for more information or file multiple applications~\cite{schwartz_hsu_cohen_2020,bhardwaj_2020}.
Although our analysis excludes those who use traditional methods for filing unemployment claims (e.g., mail, phone), our results indicate that web search is a critical resource that facilitates unemployment needs and highlight a potential gap in satisfying unemployment needs which requires further investigation. 

\xhdr{Expressed Domestic Violence Needs vs. Reports}
\squeeze{One of the dire consequences of the COVID-19 pandemic is an increased risk of domestic violence due to shelter-in-place mandates exacerbated by physical, financial, and social-emotional stressors and increased alcohol consumption at home~\cite{campbell2020increasing,li_schwartzapfel_2020}.
Our analysis shows that the expression of needs for \emph{domestic violence queries} has dropped by nearly 36.7\% (=$2^{-0.66}$--1) since the pandemic and stabilized at around -15.9\% (=$2^{-0.25}$--1, Fig.~\ref{fig:unemployment}B) below pre-pandemic levels. 
Interests in \emph{firearm purchase} have increased by over 40\% (=$2^{0.5}$--1), corroborating other reports of increased gun sales and gun violence and worrisome links to fatal domestic violence incidents~\cite{sutherland2020gun,hatchimonji2020trauma,campbell2020increasing}.}

\squeezetiny{A similar decrease in domestic violence related metrics in March is seen by national hotlines\footnote{\url{https://crisistrends.org/}; \url{https://www.thehotline.org/}}. Some reports point to potential underreporting by victims unable to reach out for help under constant surveillance by the offenders at home and fearful of exposure to the virus at a shelter~\cite{bullinger2020covid, southall_2020}. 
Although the expansion of online resources are critical to addressing the rise in domestic violence~\cite{mahase2020covid}, many studies highlight the need to recognize control tactics that prevent access to these digital resources~\cite{ross2020if}.
The sustained decrease in domestic violence queries seen in our data may indicate such barriers or shifts in access methods for these online resources.
The complexities of domestic violence are evident in the inconsistencies across many data sources, resource media, and contexts.
Our results frame an urgent question that needs to be resolved through expertise in domestic violence and social work organizations and through careful combination of multiple data sources to identify the underlying explanations for our findings.}
\section{Discussion and Conclusion}

\enlargethispage{\baselineskip}

We presented a computational methodology based on theories of human needs to quantify the effects of the pandemic and its related events on web search interactions.
Although all need categories had elevated changes in needs, we found that basic need subcategories were elevated the most while growth-based need subcategories, indicative of positive outlooks in life, were subdued (Sections~\ref{sec:tempchange},~\ref{sec:subneeds}). 
We also found that earlier and longer shelter-in-place mandates may come with an unintended impact on mental health needs (Sec.~\ref{sec:geo}).
We used unemployment and domestic violence related queries to demonstrate how our methodology could help expose gaps between the expression and fulfillment of needs--and frame questions and directions for urgent investigation (Sec.~\ref{sec:consequences}). 

\xhdr{Limitations}
We cannot use our method to make causal claims. 
If other major, concurrent events (e.g., Black Lives Matter) have significantly influenced human needs during the observations window, we are not be able to distinguish between needs caused by the pandemic or these events.
However, our analysis clearly presents changes in the expression of human needs which can be important on its own for understanding rising needs and in considering factors for the design of specific interventions.
Our analysis relies on heuristics and correct inference of needs from search interactions. But our framework is theory-driven, comprehensive, and achieves 97\% precision across five main need categories. Further fine-tuning of the framework should be motivated by precision-recall requirements of individual analysis goals. 
Although \emph{expressions} of needs may not reflect the actual underlying human needs, the methodology serves as a useful detector for needs, as we have demonstrated through its corroboration with several external findings, and combining our methodology with additional data sources and interdisciplinary collaboration can further the understanding of changes in human needs during global crises.
In contrast to traditional methods of measuring needs, our approach can operate at population scales relying solely on data that is already routinely collected by web search engines, and thus allows for effective, retrospective, and fine-grained studies with observation periods of over one year.

\xhdr{Implications: Resilience and Vulnerability}
Our work has demonstrated major shifts in a spectrum of human needs during the pandemic, echoing the need to understand the societal, economic, and psychosocial effects of these events and the underlying system of human needs. 
Such understanding is essential for guiding programs around outreach and support of people during the current pandemic and preparing for future pandemic responses and recovery efforts~\cite{ryan2020covid}. 
We observed that some needs have returned to their baseline levels while others are sustained at elevated (e.g., health and economic instability) or suppressed (e.g., job search, educational degree) levels (Fig.~\ref{fig:subneeds}). 
As we saw in Sec.~\ref{sec:tempchange}, temporal variations on how and when certain needs arise in response to major disruptions could indicate a degree of psychological and economic resilience~\cite{martin2015notion,fletcher2013psychological}.
Understanding how well a region can endure disruptions is crucial for balancing health risks against societal risks, especially for vulnerable populations who are severely impacted by the pandemic~\cite{chowkwanyun2020racial}. 
Our framework could be used to quantify how much a community has been and might be able to endure social and economic distress.
For example, a region may be able to withstand prolonged business closures before loan forgiveness needs surge, while another region may show immediate need for financial stimulus. 
In addition, when coupled with demographic variables, our framework could be used to examine disparities in needs which may help with understandings of the disparate impacts of blanket policies on the most vulnerable populations.

\xhdr{Implications: Preparedness and Resources}
Our work revealed strong correlations indicative of a long-term mental health impact of longer shelter-in-place mandate, suggesting that states with longer duration of shelter-in-place mandates may need to provide additional support for social emotional well-being (Fig.~\ref{fig:stateshelter}).
We also saw potential barriers to accessing critical resources for unemployment and domestic violence; a decrease in domestic violence queries in spite of conflicting anecdotes should raise questions and an alarm (Fig.~\ref{fig:unemployment}).
These examples suggest that expressions of needs through web interactions could highlight resource deficiencies or barriers.
As regions prepare recovery efforts from COVID-19 or make plans for future disruptions, our methodology can be harnessed to monitor or anticipate a spectrum of needs at various geotemporal granularities; this could help reveal impacts of the pandemic (e.g., mental health~\cite{cdcmentalhealth}) and cater the appropriate interventions that meet individual and societal needs.

\xhdr{Future Directions}
We see this work as a step towards achieving a more holistic understanding of the multiple influences of the pandemic and associated policies and programs. 
While we focused on retrospective analyses, our approach has the potential to be used in near real-time to monitor human needs and support future policy decisions. 
Understanding the disparate impacts of the pandemic and its policies on a full spectrum of human needs, especially for vulnerable populations, is critical for designing response and recovery efforts for major disruptions.
Our work highlighted future research opportunities and calls for action and collaboration across epidemiology, economy, social sciences, risk management, and more.
We look forward to further refinement of the methods presented and hope the work will encourage discussions on observing and addressing changes across a broad spectrum of human needs during a global crisis.

\enlargethispage{\baselineskip}

\begin{acks}
We thank the reviewers and colleagues for their valuable feedback. This research was supported in part by NSF grant IIS-1901386, Bill \& Melinda Gates Foundation (INV-004841), and the Allen Institute for Artificial Intelligence.
\end{acks}

\enlargethispage{\baselineskip}
\bibliographystyle{ACM-Reference-Format}
\bibliography{_references.bib}

\clearpage
\appendix

\setcounter{table}{0}
\renewcommand{\thetable}{A\arabic{table}}
\setcounter{figure}{0}
\renewcommand{\thefigure}{A\arabic{figure}}

\section{Online Appendix}
\subsection{Analyzing National Representation}\label{app:datademographics}
To understand how much of the US population is represented by the collected data, we obtained demographics data from the Census Reporter API~\cite{census_reporter} for ZIP codes represented in our dataset.
Census Reporter API provides demographics data for 32,989 US ZIP codes, and not all of the ZIP codes in our dataset have available demographics information through this service.
Between our dataset and the demographics data, we have 96.4\% overlap of ZIP codes, representing 97.5\% of total queries in our dataset.
Table~\ref{tab:suppdemographic} summarizes the median of the select 11 demographic variables for ZIP codes in our dataset in comparison to all available ZIP codes in Census Reporter. Given 96.4\% overlap, we find that the ZIP codes in our dataset closely mirrors the US population. 
Although query volumes are not uniformly distributed across these ZIP codes, the vast majority of US ZIP codes is included in our dataset.

\suppdemographic

\subsection{Analyzing Selection Bias}\label{app:clientrate}
We sought to understand potential biases in socioeconomic circumstances that would influence the usage of Bing search engine.
We leveraged deidentified client id as a proxy for a unique user to estimate the `client rate', or how much of the population in a ZIP code is using the Bing search engine.
We examined the correlation between the client rate and various demographic factors at the ZIP code level (e.g., income, race, age, gender, education, housing, internet access).
Although unsurprisingly the factors describe some of the variance, none were correlated more strongly than $r$=-0.058 (\% Housing Owned).
Table~\ref{tab:suppdemcorr} summarizes Pearson correlation statistics. 
These results suggest that our dataset is not strongly biased towards any single demographic.

\suppdemcorr

\subsection{Reliability of Search Interaction Trends}\label{app:googletrends}

Although our analysis relies solely on Bing search data, many Americans use other search engines such as Google. 
Therefore, we compared search trends for Bing with data available via the Google Trends API\footnote{\url{https://trends.google.com/}} for the same time period and a subset of keywords.
We chose to use select keywords from each need category, rather than applying our needs aggregation pipeline (described in Sec.~\ref{method:processing}), because the Google Trends public API does not support regular expressions or access to the click interactions.
We chose 9 keywords that were representative of their respective need subcategory (i.e., `online learning' captures \emph{online education queries}), had a significantly larger query volume compared to other keywords in the need categories (i.e., `hand sanitizer' had more query volume than `mask'), or may have a seasonal effect (i.e., query volume for `tax' could depend on the tax season). 
We conducted a correlation analysis on a moving average of a full week to account for timezone differences between the two data sources.
Visual inspection of both Google and Bing trends confirm that search patterns across these two search engines are remarkably similar (Fig.~\ref{fig:binggoogle}), and Pearson correlation coefficients are very high with a median of 0.96 (min=0.45, max=0.98, all $p$<0.001). 
Table~\ref{tab:suppbinggooglecorr} summarizes Pearson correlation outputs.
These results imply that our findings are not simply an artifact of using one search engine over another.

\suppbinggoogle
\suppbinggooglecorr

\subsection{Human Needs Categories and Detection}\label{app:needs}

We draw inspiration from Maslow's hierarchy of fundamental human needs \cite{maslow1943theory,maslow1970new} to tag each search interaction. Of the eight top-level human needs from Maslow's expanded hierarchy of needs, we omit \emph{Esteem}, \emph{Aesthetics}, and \emph{Transcendence} because they are difficult to operationalize from observational data alone.
We categorize search behaviors into one or more of five broad categories of needs. \emph{Safety} and \emph{Physiological} are considered as `basic' needs. \emph{Love and Belonging}, \emph{Cognitive}, and \emph{Self Actualization} are often considered to be `psychological' needs, and \emph{Cognitive} and \emph{Self Actualization} are considered as `growth' needs. 

We obtain further granularity of the 5 human need categories by decomposing each need category into many sub-need categories. 
This extra level of granularity has advantages in teasing apart the specific aspects of the need (e.g., shelter vs. finances in \emph{Safety} needs) and in debugging and refining the detection logic. Table~\ref{tab:suppsubneeds} enumerates 79 sub-needs and their examples. 

For each need subcategory, we had multiple regular expressions applied to either the query string, the clicked URL, or both. This is noted in Table~\ref{tab:suppsubneeds} under the `Logic' column. `Keyword and domain (KD)' logic indicates that there is one regular expression for matching query strings and another regular expression for matching clicked URLs. Both must be matched for the search interaction to be categorized as that need. `Queries (Q)' logic indicates that there is one regular expression for query strings alone regardless of the clicked URLs. `Domains (D)' logic indicates that there is one regular expression for clicked URLs regardless of the query strings. A full table of regular expressions used in the data collection is provided in a separate supplementary file and can be cross-referenced through the `Need Id' column\footnote{\url{https://github.com/jinasuh/pandemic_needs}}.

There were several steps in arriving at these subcategories and regular expressions. 
We first identified a set of e-commerce websites (1) from market research blog posts and reports that were publicly available online and (2) from enumerating top 100 URL hosts from a sample of search interactions that were tagged as purchase or commerce related using built-in classifiers in Bing. 
We combined these two data sources to manually curate a list of 61 e-commerce domains. 

Next, we collected another sample of query strings that led to subsequent clicks to these e-commerce domains as well as page snippets that were displayed on the search result page. 
Using the Latent Dirichlet Allocation module in gensim, a popular topic modeling tool in python, we trained an unsupervised topic model on the query strings and the associated page snippets for 100, 500, and 1000 topics on 100,000 unigrams. 
We manually inspected the keywords in the topics to curate 22 topics of over 800 keywords. 
We also extracted frequent bigrams from the same dataset to expand our keywords if existing unigrams were too ambiguous.
These topics represented purchase categories because they were collected from search interactions with subsequent clicks to e-commerce websites. 

Our next step was to categorize these topics into five main need categories. 
We used a hybrid card sorting method~\cite{albert2013measuring} with 5 participants who merged, split, edited, or created these topics into need subcategories. Then these need subcategories were then placed into the 5 main need categories. 

Using the outputs from the card sorting activity as a basis, we further brainstormed subcategories beyond purchase categories. 
Several researchers independently brainstormed and categorized subcategories which were combined and resolved collaboratively through consensus meetings.
We referred to the definitions presented in the theories of human needs to resolve any remaining disagreements~\cite{maslow1943theory}. 

Once the need categories and subcategories were identified, we leveraged keywords and domains from manually curated topics and card sort outputs.
We also obtained a list of topical domains from publicly available blog posts and articles summarizing recommended websites (e.g., search for best parenting website) or manually curated a list of domains (e.g., search for unemployment benefit page for each state).
Again, we independently brainstormed and curated lists of keywords, keyword patterns, and domains for each subcategory and collaboratively iterated on the lists through consensus meetings.
These keywords, keyword patterns, and domain URLs were combined into many regular expressions that we used to tag each search interaction.

We validated our detection logic by collecting human labels from Amazon's Mechanical Turk, and our evaluation set achieved a precision of 97.2\%, as described in detail in Sec.~\ref{sec:validation}. 
A screenshot of an example task seen by a crowd worker is shown in Fig.~\ref{fig:suppmturktask}, and the full set of detailed task instructions can be found at the end of this Appendix.

\suppsubneeds
\suppfigmturktask

\newpage

\subsection{Significant Changes in Human Needs}\label{app:subneeds}

\suppfigsubneeds

\newpage


\section*{Supplementary material}

# Human Need Detection Instructions

A search query can be an expression of an underlying human need.

For example, if I am running a fever and want to purchase a fever reducer online (physiological need), my query might be "tylenol" with a subsequent click to amazon.com in the search results to purchase the item. If I am pursuing baking as a hobby (self actualization need), my query might be "how to bake a perfect sourdough", and the subsequent clicks do not matter much to understand the intent for that query. Or if I am visiting coursera.org, it might indicate that I want to learn a new topic (cognitive need).

In this task, please read the search query and the subsequent website visit (if available) to determine which of the following human needs category it belongs to. Search queries should seek information, express interest, intend to purchase or obtain goods that could satisfy these needs.

# Human Needs Definitions

## Self Actualization Need

Self actualization needs are about realizing personal potential, seeking personal growth, and self-fulfillment. Some of the topics to consider are:

- Recreation or hobbies such as reading, technology, craft, music, art, pets, photography, media, sports, toys, cooking, home improvement, outdoor
- Parenting and child rearing
- Marriage or wedding
- Talent acquisition
- Life or personal goals
- Charity, donations, volunteering

> *Example 1:*
> 
> **Search Query:** how to photograph birds in flight
> 
> *Example 2:*
> 
> **Search Query:** wedding invitation design
> **Visited Url:** theknot.com

## Cognitive Need

Cognitive needs are about pursuing knowledge and intelligence through learning and discovering. Some of the topics to consider are:

- Education or learning materials

- Online or virtual classroom
- Educational degrees or programs
- Cognition, memory, focus or attention

> *Example 1:*
>
> **Search Query:** lesson plans for 4th grade math
>
> *Example 2:*
>
> **Search Query:** python data science
> **Visited Url:** coursera.org

## Love and Belonging Need

Love and belong needs are social and emotional and include emotionally-based relationships such as friendship, family, dating, sexual intimacy. Some of the topics to consider are:

- Expression of or resources for mental health or emotional issues such as anxiety, depression, loneliness, isolation, suicide, nervousness, rejection, fear or sadness
- Social media, social network and relevant technologies
- Social activities online or offline
- Search for relationships, significant others, dating
- Negative relationships such as divorce or breakup

> *Example 1:*
>
> **Search Query:** online therapist for anxiety
> **Visited Url:** talkspace.com
>
> *Example 2:*
>
> **Search Query:** long distance relationships

## Safety Need

Safety needs stem from our desire to seek order and stability in the world. These needs provide protection from elements through shelter and security of body, jobs, food, resources, family, or health. Some of the topics to consider are:

- Protection from harm through equipments such as firearms or security systems
- Personal protective equipment such as masks or sanitizers
- Lack of protection such as domestic violence

- Financial related such as tax, loans, banking, or bankrupsy
- Jobs and unemployment
- Shelter related such as housing, rental, evictions
- Assistance for jobs or finances

> ***Example 1:***
>
> **Search Query:** n-95 masks
> **Visited Url:** amazon.com
>
> ***Example 2:***
>
> **Search Query:** unemployment benefits

## Physiological Need

Physiological needs are the basic animal needs such as air, food, drink, shelter, warmth, sex, and other body needs. Some of the topics to consider are:

- Health related such as prescription, measurement, symptoms, conditions, or first aid
- Hygienic products such as toilet paper
- Basic staples such as food, beverages, apparel, household products
- Services that provice basic staples like grocery delivery or online grocers
- Sleep and insomnia
- Assistance for food
- Transportation and mobility

> ***Example 1:***
>
> **Search Query:** Symptoms of flu
>
> ***Example 2:***
>
> **Search Query:** whole foods
> **Visited Url:** wholefoodsmarket.com


\end{document}